# Surface Structural Disordering in Graphite upon Lithium Intercalation/Deintercalation

**Vijay A. Sethuraman, Laurence J. Hardwick,[1] Venkat Srinivasan and Robert Kostecki[*]**

Environmental Energy Technologies Division
Lawrence Berkeley National Laboratory
Berkeley, CA 94720-8168, USA

We report on the origin of the surface structural disordering in graphite anodes induced by lithium intercalation and deintercalation processes. Average Raman spectra of graphitic anodes reveal that cycling at potentials that correspond to low lithium concentrations in $Li_xC$ ($0 \leq x < 0.16$) is responsible for most of the structural damage observed at the graphite surface. The extent of surface structural disorder in graphite is significantly reduced for the anodes that were cycled at potentials where stage-1 and stage-2 compounds ($x > 0.33$) are present. Electrochemical impedance spectra show larger interfacial impedance for the electrodes that were fully delithiated during cycling as compared to electrodes that were cycled at lower potentials ($U < 0.15$ V *vs.* $Li/Li^+$). Steep $Li^+$ surface-bulk concentration gradients at the surface of graphite during early stages of intercalation processes, and the inherent increase of the $Li_xC$ d-spacing tend to induce local stresses at the edges of graphene layers, and lead to the breakage of C-C bonds. The exposed graphite edge sites react with the electrolyte to (re)form the SEI layer, which leads to gradual degradation of the graphite anode, and causes reversible capacity loss in a lithium-ion battery.

[1]Present address: School of Chemistry, University of St Andrews, North Haugh, St Andrews, KY16 9SS, Scotland, United Kingdom.
[*]Corresponding author:
Tel: +1 (510) 486-6002; fax: +1 (510) 486-7303; email: R_Kostecki@lbl.gov

## 1. INTRODUCTION

Understanding the mechanisms of aging processes and degradation modes of lithium-ion systems remains an important objective for battery research [1,2]. This is most notably true for lithium-ion batteries for transportation applications, where 10–15 years of battery lifetime is required [3]. Though several different anode chemistries are currently pursued, graphite is still the primary choice for anodes used in commercial lithium-ion batteries [4,5]. However, graphitic anodes suffer severe surface structural disordering upon prolonged cycling in rechargeable lithium-ion batteries [6,7,8,9,10]. This deleterious effect is intensified at high charging rates and elevated temperatures as evidenced in the Raman spectra of graphite anodes sampled from aged/cycled lithium-ion cells, which show an increased intensity of the carbon D-band (*ca.* 1350 cm$^{-1}$) with respect to the G-band (*ca.* 1580 cm$^{-1}$) [11,12,13]. This surface structural disorder is continuously inflicted on the graphitic crystallites in the anode upon prolonged charge/discharge cycling, modifies their electrocatalytic properties, and consequently, affects the thickness and composition of the solid-electrolyte-interphase (SEI) layer [14,15,16]. The continuous reduction of the electrolyte and reformation of the SEI layer results in the gradual loss of cyclable lithium and consumption of the electrolyte. Since lithium is a finite resource in a typical lithium-ion battery, this loss is directly responsible for its capacity fade, and eventual failure.

The objective of this work is to investigate the origin of the surface structural disordering in graphite, and the relationship between the amount of surface structural damage, cycling conditions, and the electrochemical performance of graphitic anodes in Li-ion battery systems.

## 2. EXPERIMENTAL

*2. 1. Electrode and coin cell fabrication*

Composite anodes [Mag-10, Hitachi, 92%, poly(vinylidene fluoride (PVDF) 8%] were produced from *N*-methyl pyrrolidinone (Sigma Aldrich) slurry coated onto a Cu-foil (thickness = 0.3 mm). The anodes (disc Φ = 1.2 cm) were dried at 120°C under vacuum for 24 hours and then transferred into an Ar-filled glove box (Nexus II, Vacuum Atmospheres Co.) without exposing them to ambient air. The anodes were then assembled into sealed 2325-type coin cells with a Li-foil counter and reference electrodes, Celgard 2500 separator (Celgard Inc.) soaked with 1.2 M lithium hexafluorophosphate in a mixture of ethylene carbonate and diethyl carbonate [1.2 M LiPF$_6$ in EC:DEC (1:2, by % wt.) electrolyte (Ferro Corp.)].

*2. 2. Electrochemical measurements*

All electrochemical measurements were conducted in a temperature chamber (Test Equity) at 23°C (±1°C) using a 1480A MultiStat system (Solartron Analytical) furnished with Corrware (Scribner Associates Inc.) and Z-plot (Southern Pines) software. The galvanostatic formation cycles at $i$ = 0.15 mA cm$^{-2}$, which corresponds to a charge/discharge rate of C/25 rate (C = 372 mAh g$^{-1}$ denotes the theoretical charge capacity of the carbon electrode and C/25 corresponds to a current allowing a full discharge in 25 hr), were conducted from the open circuit potential (*ca.* 3 V *vs.* Li/Li$^+$, first scan) to 0.01 V *vs.* Li/Li$^+$ and then between 1 V and 0.01 V *vs.* Li/Li$^+$ for a total of three cycles. All potentials in this study are referred to the Li/Li$^+$ reference electrode.

Long-term cycling experiments were carried out galvanostatically at C/5 rate between the following three potential limits: 1 – 0.18 V *vs.* Li/Li$^+$, 0.23 – 0.098 V *vs.* Li/Li$^+$ and 0.015 – 0.005 V *vs.* Li/Li$^+$, which correspond to *ca.* Li$_{0 \leq x < 0.1}$C, Li$_{0.1 < x < 0.5}$C, and Li$_{0.3 < x \leq 1}$C compositions, respectively. These potential limits were chosen because they allow for the cells to be cycled distinctly between dilute, intermediate and concentrated stages. Note that the potential offset between the intercalation and deintercalation curves were also taken into account for arriving at these cycling limits. A total of 200 cycles were carried out for each cell. The cells corresponding to these cycling protocols are referred to as A1, A2 and A3, respectively. Thus the A1 cell was cycled between pure graphite and dilute stage-4, A2 cell was cycled between stage-4 and stage-2 compositions, and A3 cell was cycled between stage-3 and stage-1 compositions. Electrochemical impedance spectra were recorded every 50 cycles at open circuit potential on a completely discharged cell within 10 mHz to 1 MHz frequency range (1252A Frequency Response Analyzer, Solartron Analytical). Note that the EIS experiments were conducted on a two-electrode cell, and the contribution from the change in the impedance of the lithium-metal reference and counter electrode was ignored because (a) the currents during cycling were small (*i.e.,* C/5 rate) and (b) excess lithium was used.

*2. 3. Raman spectroscopy measurements*

After cycling, the electrodes were removed from the coin cells, washed with dry dimethyl carbonate (DMC) to remove any remaining residual EC, $LiPF_6$, and left to dry in the glove box for 10 minutes. The electrodes were placed in an air-tight spectroscopic cell equipped with a glass optical window (thickness = 0.15 mm) and were analyzed by Raman microscopy (Labram, ISA Groupe Horiba) with a HeNe laser ($\lambda$ = 632.8 nm, 1 mW power) as the excitation source. Four 48 x 74 μm Raman maps from different locations at the electrode surface of were collected in autofocus mode with a spatial resolution of *ca.* 0.7 μm. The average Raman spectra as well as the average D-band to G-band intensity ratio ($I_D/I_G$) for each electrode were calculated from all the spectra in Raman maps.

**3. RESULTS AND DISCUSSION**

Figure 1 shows a typical potential profile of the lithium-ion intercalation/deintercalation in the Mag-10 graphite electrode at C/25 rate. The distinct plateaus seen in both the intercalation and the deintercalation potential profiles correspond to stage 1-4 compounds [17] as indicated by the schematic shown above based on the Daumas-Hérold model [18]. The potential plateau at *ca.* 210 mV corresponds to the transition to a stage 3 graphite-intercalation compound (GIC) from a dilute stage 1 *via* stage 4, the potential plateau at *ca.* 160 mV *vs.* $Li/Li^+$ corresponds to a transition from stage-3 GIC to a stage-2 liquid phase (*i.e.,* no in-plane ordering), the potential plateau at *ca.* 130 mV *vs.* $Li/Li^+$ corresponds to a transition between stage-2 liquid phase to an ordered stage-2 GIC, and the potential plateau at *ca.* 80 mV *vs.* $Li/Li^+$ corresponds to a transition from an ordered stage-2 GIC to a stage-1 GIC [4,17]. Note that the schematic above the charge-discharge curve represents the system in equilibrium, and not during cycling at a C/25 rate.

Representative average Raman spectra (averaged from 5400 points) of fresh and cycled electrodes are shown in Figure 2. The D-band observed at 1350 $cm^{-1}$ corresponds to the $A_{1g}$ vibrational mode [19] and can be attributed to the breathing motion of $sp^2$ hybridized carbon atoms in rings at edge planes and defects in the graphene sheet [20]. The G-band observed at 1580 $cm^{-1}$ corresponds to the $E_{2g}$ vibrational mode and is due to the relative motion of $sp^2$ carbon atoms in rings as well as chains. The peak intensity ratio $I_D/I_G$ is often used to determine the extent of structural disorder (*e.g.*, $I_D/I_G = 0$ for a perfect, infinite graphene layer) in graphite and/or the size of the graphitic domains [21,22].

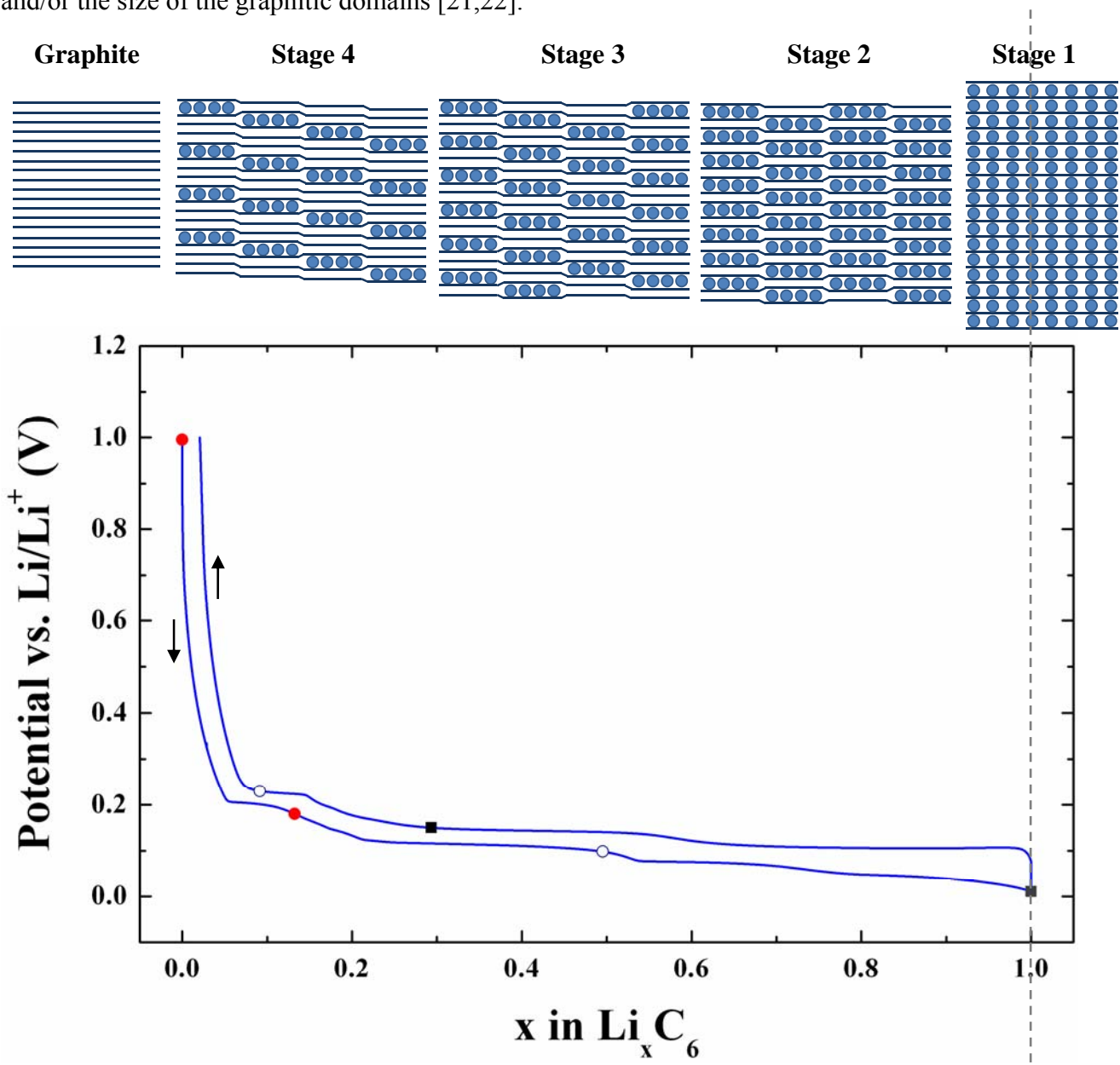

*Figure 1: Charge-discharge profile of a MAG-10 electrode at ca. C/25 rate shown vs. composition along with the upper- and lower-potential limits (shown as symbols) for the three different cycling protocols: 1.0 and 0.18 V vs. Li/Li$^+$ (●), 0.098 and 0.23 V vs. Li/Li$^+$ (○) and 0.005 and 0.15 V vs. Li/Li$^+$ (■). Schematic representation of the staging phenomena is shown above the charge-discharge curve.*

The Raman spectra results are summarized in Table 1 along with the potential, and the composition limits. The relative D- and G-peak heights change noticeably for the cycled anodes.

The D-band intensity increases substantially, the G-band broadens slightly. The relative average intensity ratios of the D- and the G-bands increased from 0.25 for pristine electrode to 0.45, 0.58 and 0.61 for the cycled anodes, respectively, indicating that severe structural damage was induced into the graphite during cycling. The A1 cell, which sustained only shallow $Li^+$ intercalation/deintercalation, suffered the most damage to the graphite surface structure. The graphite electrode from the cell A2 that was cycled at the higher $Li^+$ concentration ranges in graphite display noticeably less structural degradation. Interestingly, the least affected electrode originates from the A3 cell that was cycled very close to the fully lithiated $LiC_6$ state.

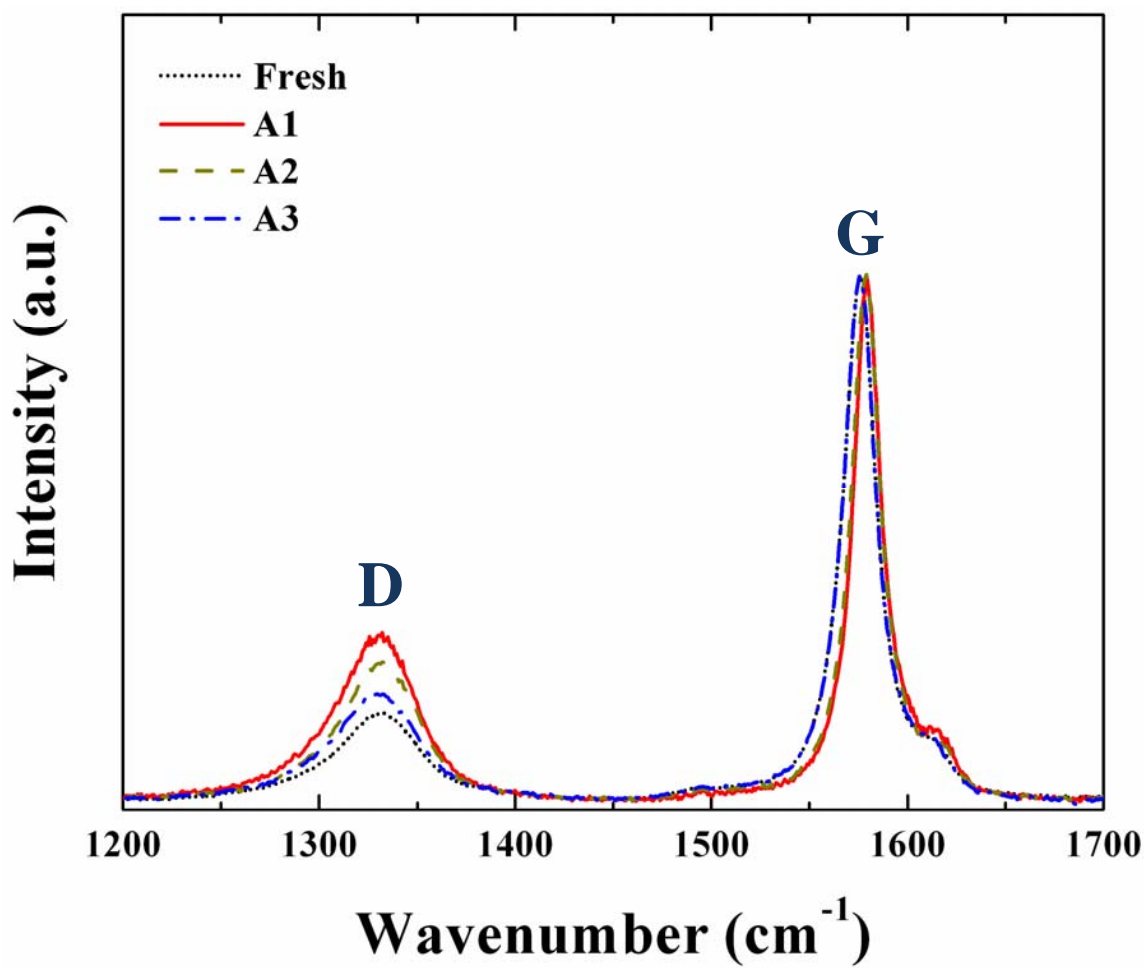

*Figure 2: Average Raman spectra of fresh and cycled electrodes. Band intensities are normalized to the G-band.*

Representative Raman $I_D/I_G$ ratio surface maps obtained from (a) pristine MAG-10 electrode and (b) MAG-10 electrode cycled between 1 and 0.18 V are shown in Figure 3. The $I_D/I_G$ ratios were derived from each individual Raman spectra recorded in the mapping area at 0.7 µm spatial resolution. Dark areas on the map correspond to highly graphitic carbon with low $I_D/I_G$ ratios, whereas light areas represent disordered graphite with elevated $I_D/I_G$ ratios.

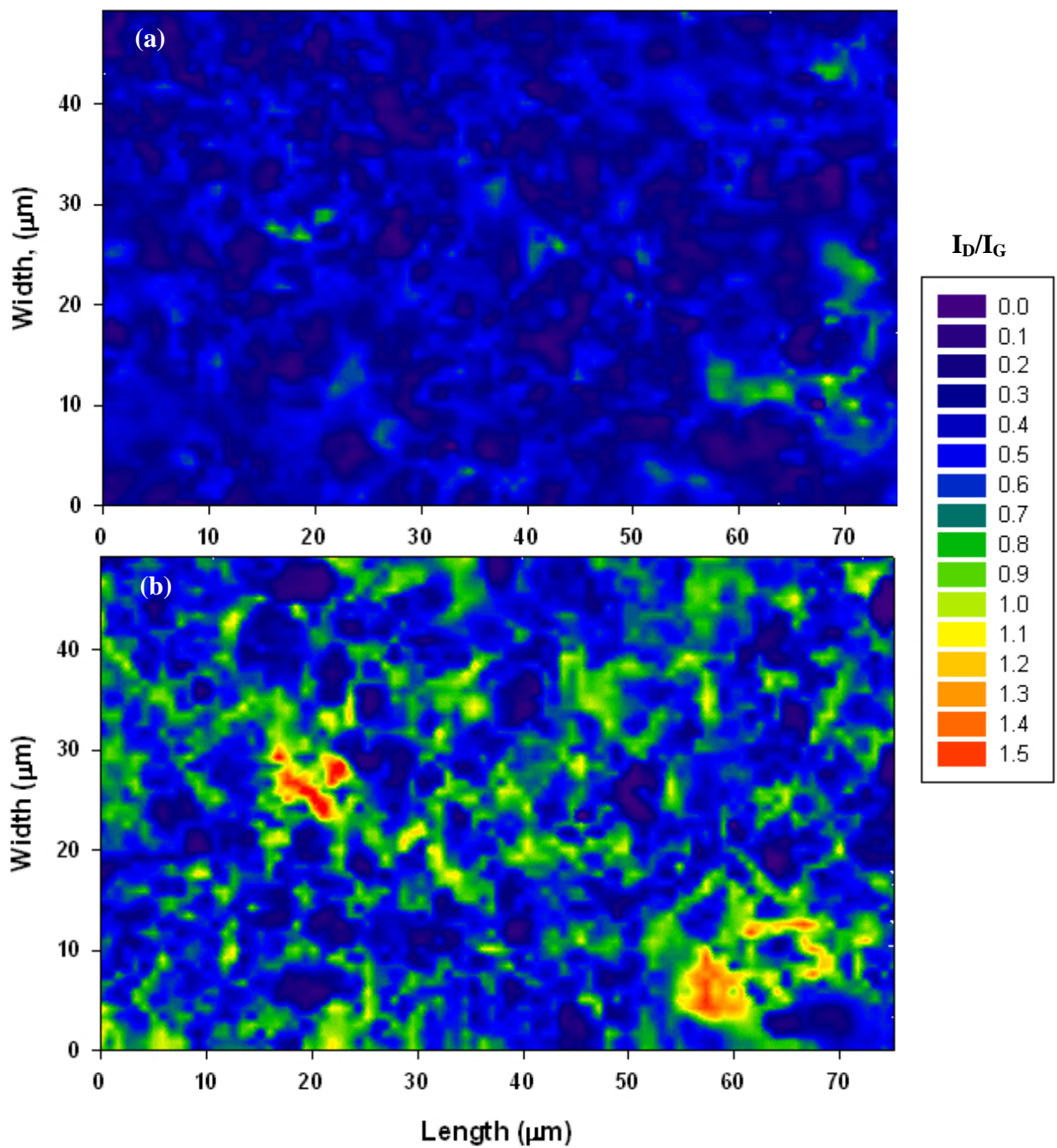

Figure 3: Surface Raman maps of the $I_D/I_G$ ratio from 48 x 74 µm area at ca. 0.7 µm resolution of (a) pristine graphite electrode, and (b) electrode cycled between 1V and 0.18 V vs. $Li/Li^+$.

The fresh anode displays a fairly uniform graphitic structure with some local disorder. The increased prevalence of dark areas at the surface of the cycled graphitic anode indicates an increased extent of local graphite structural degradation. Some severe local structural disorder is observed in the cycled anode with only a few local areas which retained the original graphitic structure. This Raman surface map of the cycled anode shows clearly that graphite structural degradation proceeds in a highly non-uniform manner.

The electrochemical-impedance-spectroscopy results are summarized in Figures 4a-d. The low frequency tail of the spectra (*i.e.,* 30-40 $\Omega$ cm$^{-2}$ range) that corresponds to Li$^+$ diffusion in the electrolyte and solid-state diffusion of lithium-ion into graphite remains relatively unaffected. The mid-low frequency semi-circle consists of several contributing factors that are associated with charge-transfer resistance [23], ohmic contact resistances between graphite particles [9,24], Li$^+$ transport across the SEI layer, and electron transfer between graphene sheets [25]. Small variation in high-frequency resistance values (*i.e.,* 3-5 $\Omega$ cm$^{-2}$ range) exists between the three cells, possibly due to fabrication. All three cells show a noticeable increase in the mid-frequency section of the impedance spectra upon cycling. This could be due to a buildup of significant mass-transfer and charge-transfer barriers across the SEI layer at the surface of graphite particles during long term cycling. Though subtle, the observed impedance increase is the highest for A1 cell followed by A2 and A3 cells. This impedance behavior pattern corresponds exactly to the extent of surface carbon disordering observed by the Raman measurements. This is in concert with our earlier studies, which have shown that the surface disordering of the graphite upon cycling results in the continuous reformation of SEI, leading to a thicker SEI layer, and consequently, higher interfacial resistance [9].

Lithium-ion intercalation and deintercalation in graphite occurs *via* staging, *i.e.,* formation of metastable phases, which are defined by the energy required for the guest species (*i.e.,* Li$^+$) to break the van der Waals interactions between host graphene layers, and the repulsive interactions between guest species. The electrochemical intercalation of Li$^+$ in graphite proceeds from a dilute stage-4 (*ca.* Li$_{0.05}$C at 0.2 V *vs.* Li/Li$^+$) to a concentrated stage-1 compound (LiC$_6$ at 0.05 V *vs.* Li/Li$^+$) [4,17,26]. Though this phenomenon is not completely understood, it is widely accepted that Li$^+$ intercalation in graphite follows the staging-domain model or the pleated-layer model proposed by Daumas and Hérold in 1969 [18]. According to the pleated-layer model, the average number of intercalating ions between any two graphene layers of the macroscopic graphite crystal is the same. The model assumes that ions can move only in between graphene layers whereas ion transport across or around graphene layers is not allowed [27].

The average space between graphene layers gradually increases upon lithium-ion intercalation from 3.359 Å for pristine graphite to 3.712 Å for stage-1 LiC$_6$ compound [28]. This rather mild structural rearrangement is not expected to rupture or displace permanently, the graphene layers within graphite crystallites. As the matter of fact, the graphene layers in a graphitic crystallite are flexible and tend to deform around the intercalating lithium ions (Figure 1) with a bending modulus value of 9.93 x 10$^{-20}$ J, as measured *via* phonon dispersion experiments [29]. This is contrary to the Rüdorff model [30], which proposes a sequential filling up of alternating graphene interlayer spaces with no structural distortions induced within the individual graphene sheets.

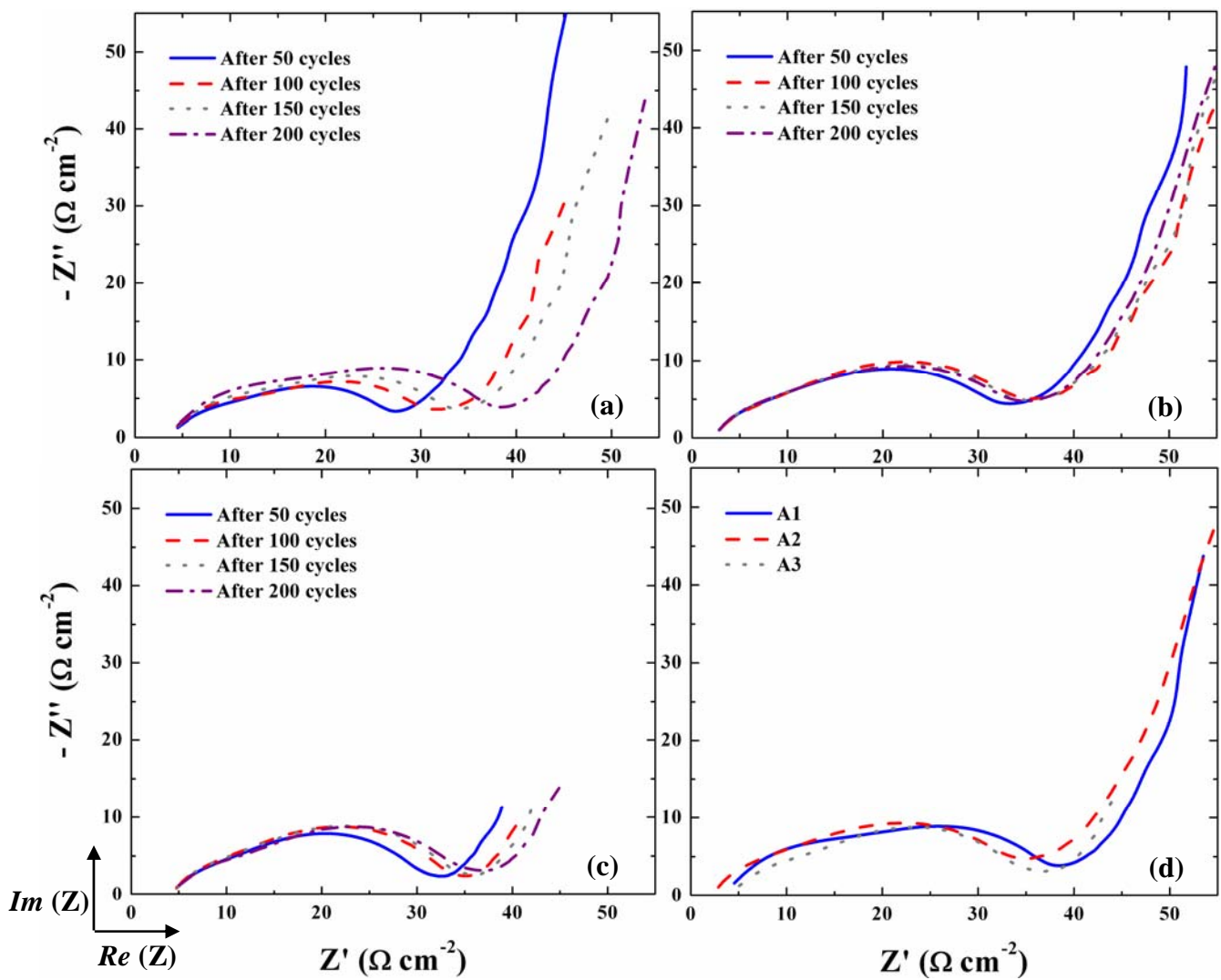

*Figure 4: Electrochemical impedance spectra recorded at every 50 cycles for A1, A2, A3 cells cycled at the three different cycling protocols: a) 1.0 and 0.18 V vs. Li/Li$^+$, b) 0.098 and 0.23 V vs. Li/Li$^+$, and c) 0.005 and 0.15 V vs. Li/Li$^+$, respectively. (d) The impedance spectra of all three cells after 200 cycles.*

Intercalation/deintercalation at high rates may create high local concentration gradients and induce local stresses within the lattice, which may eventually result in structural damage. This is particularly likely during early stages of Li$^+$ intercalation when the relatively high Li$^+$ concentration at the edge sites of graphite crystallite *vs.* empty interlayer sites in the graphite bulk will create tremendous stress at the edges of graphene sheets. This local stress may lead to a severe deformation of the graphene layers, and eventually, breaking of C-C bonds and carbon disordering. In fact, it has been shown that the larger intercalation-ions (*e.g.,* 1-ethyl-3-methylimidazolium ions) tend to induce more disorder in microcrystalline graphite than smaller ions (*e.g.,* $Li^+$) [31]. The freshly exposed carbon atom surface sites will immediately react with the electrolyte and contribute to the SEI layer.

Lithium-ion intercalation and transport in graphene crystallites involves a series of basic, surface and bulk phenomena. Theoretical (and experimental) studies indicate that $Li^+$ ions tend to form stronger bonds with carbon edge atoms than in between graphene layers [32] and electron transfer rates on edge-plane graphite are *ca.* $1 \times 10^5$ times higher than basal-plane graphite [33]. This is because most of the electrons with high energy are predominantly localized on the surface-active edge sites. Furthermore, lithium transport in graphite in highly anisotropic and lithium atoms tend to diffuse toward the edge sites where they are preferentially bound [32]. Thus, the surface concentration of $Li^+$ in graphite during intercalation/deintercalation processes is always higher than in the bulk. The resulting concentration gradient between the fully occupied surface sites and the bulk induces a significant local stress and lattice deformation in the graphene layers in the vicinity of their edges. One can expect structural stress associated with intercalating and deintercalating a graphite electrode during early phases of the $Li^+$ intercalation (*i.e.,* formation of stage-4 and stage-3 compounds), and final steps of deintercalation processes (*i.e.,* complete delithiation of $Li_xC$) (see Figure 5).

The $Li^+$ surface-bulk concentration gradient and the induced stress in the graphite lattice gradually diminish during the formation of $Li^+$-rich stage-3 or stage-2 not to mention stage-1 compounds. Therefore the observed amount of crystalline disorder generated during charge/discharge cycling between more concentrated stages (x>0.1) is significantly lower than during cycling between dilute stages (x<0.1).

These results point at the origin of one of the graphite-degradation modes in Li-ion batteries, which may have serious implications for the battery's electrochemical performance, calendar and cycle-life. It appears that shallow cycling of graphitic anodes (*i.e.,* between dilute $Li_xC$ stages and pristine graphite) should be avoided in order to minimize the surface-structural damage, the SEI layer reformation processes, impedance rise and loss of cyclable lithium in the battery. Therefore, complete discharge of commercial lithium-ion batteries should be avoided so that graphite anodes do not experience the transition between a dilute $Li_xC$ and pure graphite upon charge.

Furthermore, chemical grafting of the edge-carbon sites to weaken the strength of $C_{edge}$-$Li^+$ bonds and/or using electrolyte additives to help quickly reform the SEI at the damaged sites may be considered as strategies to minimize the observed surface-structural disordering and reduce its detrimental effects on the anode and the Li-ion system.

## 4. ACKNOWLEDGEMENTS

The authors gratefully acknowledge the financial support from the Assistant Secretary for Energy Efficiency and Renewable Energy, Office of Vehicle Technologies, the United States

Department of Energy, under contract no. DE-AC02-05CH11231. The authors thank Dr. Vincent Battaglia and Dr. Gao Liu for the provision of the electrode material and Dr. Paul Berdahl for helpful discussions.

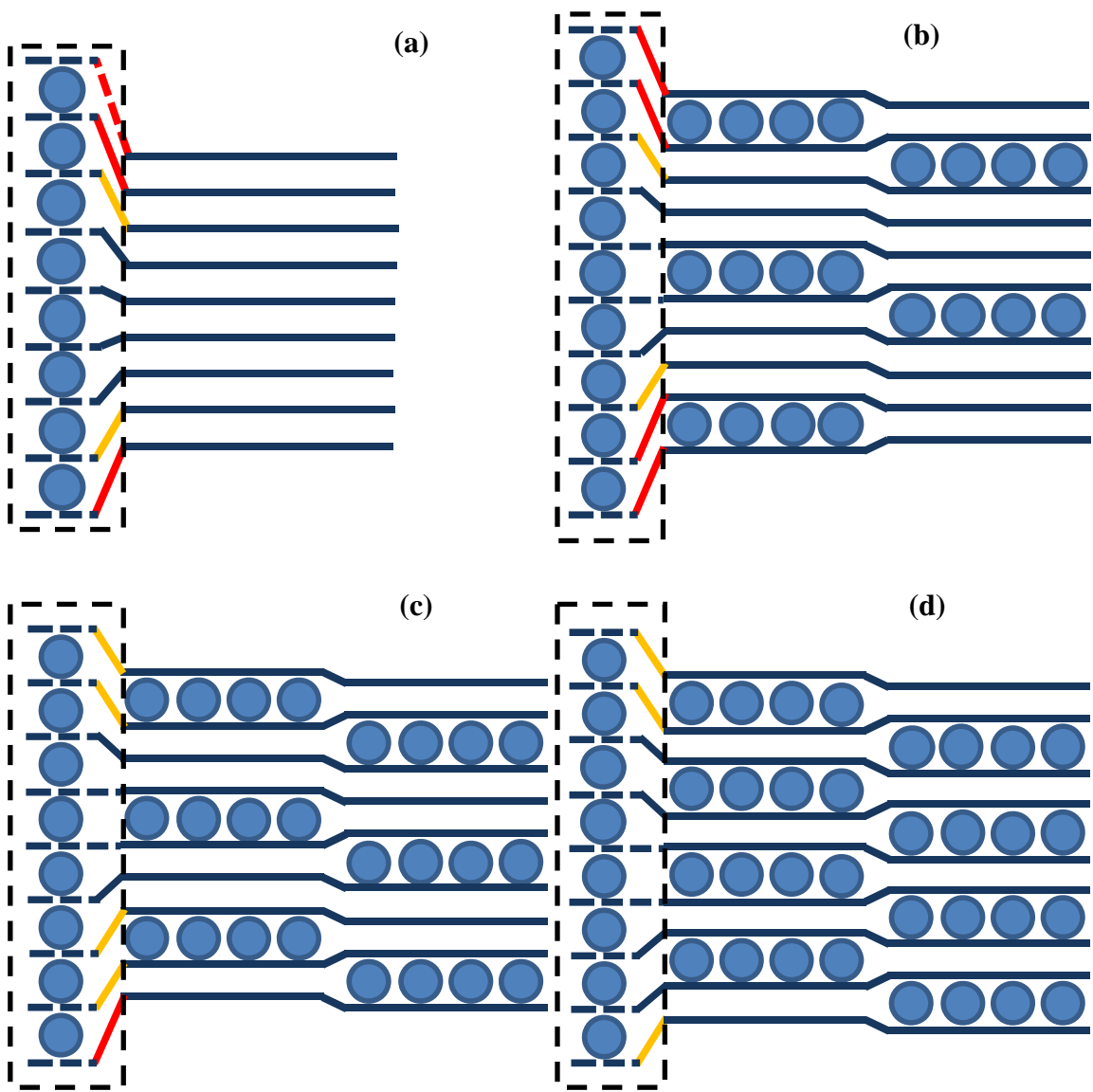

Figure 5: Schematics showing surface non-equilibrium conditions for $Li^+$ intercalation into (a) graphite, (b) stage-4, (c) stage-3 and (d) stage-2 to emphasize the influence of concentration gradients along the length of the graphene sheets as one would expect during departure from equilibrium conditions (e.g., during cycling at moderate to high rates). The larger stretching of the graphene sheet is represented in red and moderate stretching is represented in orange. The lines and the circles represent graphene sheets and lithium intercalants, respectively.